\documentclass[a4paper,prl,twocolumn, superscriptaddress]{revtex4}
\usepackage[utf8]{inputenc}
\usepackage{hyperref}
\usepackage{natbib}
\usepackage{graphicx}
\graphicspath{{figures/}}
\usepackage{siunitx, amsmath}
\begin{document}
\title{Control of ultrashort spin current pulses in metallic multilayers}
\date{\today}

\author{Alexey Melnikov}
\email{alexey.melnikov@physik.uni-halle.de}
\author{Liane Brandt}
\author{Niklas Liebing}
\author{Mirko Ribow}
\author{Ingrid Mertig}
\author{Georg Woltersdorf}

\affiliation{Institute of Physics, Martin Luther University Halle-Wittenberg, Von-Danckelmann-Platz 3, 06120 Halle, Germany}

\begin{abstract}
We study the emission of femtosecond spin current pulses at Fe/Au interfaces. Using optical second harmonic generation we demonstrate the control of pulse shape and duration by varying the emitter thickness and show that it is determined by the parameters of hot electrons emission and the diffusive transport. Using a simple model of electron emission we develop a description of superdiffusive spin transport in Au allowing to extract electron velocity and scattering rates from the experimental data. The importance of carrier multiplication in Fe is demonstrated. 
\end{abstract}

\maketitle

\begin{figure}
\centering
\includegraphics[scale=0.65]{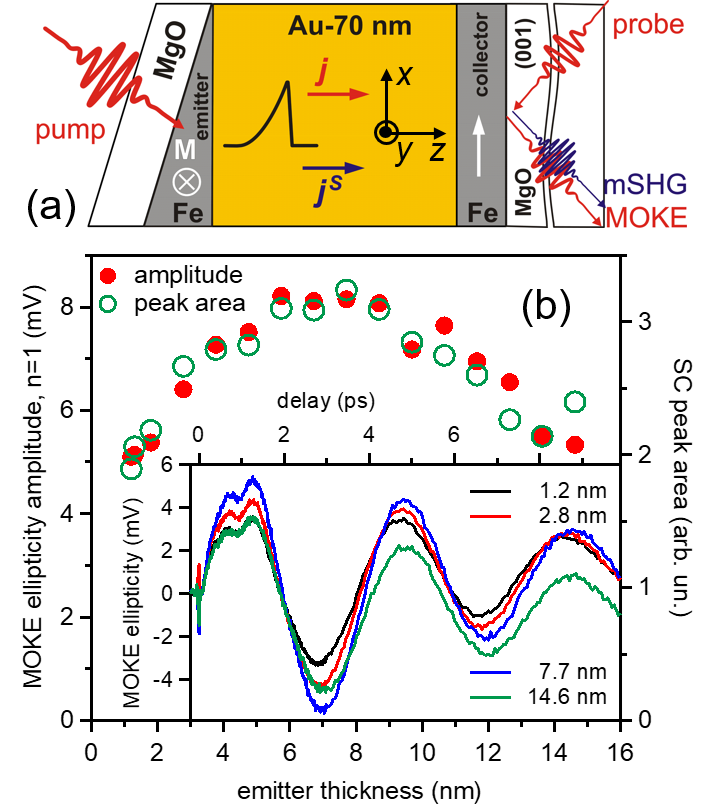}
\caption{(a) Experimental approach with the coordinate frame. (b) The amplitude of first PSSW mode measured by MOKE ellipticity along with the spin current peak area vs. the emitter thickness. The inset shows exemplary traces of MOKE ellipticity for indicated emitter thicknesses.}
\label{fig:Fig2}
\end{figure}

The observation of magnon excitation in antiferromagnets \cite{Kampfrath2011, Krieger2015,Baierl2016} and switching in ferrimagnetic materials \cite{Stanciu2007, Yang2017} has triggered enormous interest in the field of ultrafast spintronics. The further development of this field would greatly benefit from efficient techniques for generation of ultrashort spin current pulses. 
At slower time scales spin polarized currents are already used to switch the magnetization in nano-structures via the spin transfer torque mechanism \cite{Krivorotov2005}. These spin currents are typically generated either by polarizing them in metallic ferromagnets \cite{Katine2000} or using the spin Hall effect \cite{Hirsch1999,Liu2012}. For spin currents propagating across normal metal (NM)/ferromagnet (FM) bilayer interfaces, the transverse spin component (i.e. orthogonal to the FM magnetization) of spin polarized currents is absorbed within a few atomic layers at the NM/FM interface \cite{Stiles2002}. 
In recent years it became evident that spin currents also play a central role for the process of ultrafast optical demagnetization in metallic ferromagnets \cite{Beaurepaire1996, Koopmans2005,Melnikov2011,Malinowski2008,Rudolf2012}. Furthermore, it was demonstrated that the corresponding spin currents can be used to excite spin waves in the THz frequency range using the spin transfer torque mechanism \cite{Schellekens2014,Razdolski2017}. 
The optically induced emission of spin currents in FM/NM interfaces can be understood in terms of the non-thermal spin-dependent Seebeck effect (see Supplemental Material \cite{supp}, Note 1). This mechanism is operative if three conditions are fulfilled: (i) spin- and energy-dependent transmittance of FM/NM interfaces, (ii) a broad energy distribution of photo-excited hot carriers in FM, and (iii) small optical absorption of light in NM leading to large difference of hot electron concentrations in FM and NM shortly after the excitation. The generated spin current pulses with a typical duration of 250 fs, are absorbed by a magnetic collector layer (see Fig.~\ref{fig:Fig2}a) and trigger ultrafast spin dynamics \cite{Razdolski2017}. Due to the interface-confined excitation a series of perpendicular standing spin waves (PSSW) with THz frequencies is excited and can be detected by time resolved magneto-optical Kerr effect (MOKE) \cite{Razdolski2017, Brandt2021}. 

In this letter we examine the spin current generation process using the example of well defined epitaxial Fe/Au/Fe layer structures (for sample details see \cite{Brandt2021}) and optimize the spin current emitter performance in terms of amplitude and bandwidth. Using direct optical methods we study the spin current pulse shape and develop a model description linking the optical data to transport parameters.
Specifically, we vary the thickness of the spin current emitter thickness. For this, the emitting Fe layer is prepared in the form of a wedge as illustrated in Fig.~\ref{fig:Fig2}a. Optical pump-probe experiments use a back pump-front probe approach (see Supplemental Material \cite{supp}, Note 2), similar to those used in \cite{Melnikov2011,Alekhin2017,Razdolski2017,Alekhin2019,Brandt2021}. The emitter thickness $d_E$ ranges from 1 to 17~nm while the thickness $d_S$ and $d_C$ of Au spacer and Fe collector layers are fixed at 70 and 4.4~nm, respectively. The inset of Fig.~\ref{fig:Fig2}b shows exemplary traces of MOKE ellipticity excited and detected with 800~nm, 15~fs pulses. By fitting the data to exponentially damped cosine functions we obtain the excitation amplitude of the 1st PSSW mode as a function of $d_E$ as shown in Fig.~\ref{fig:Fig2}b. This quantity represents the spin transport efficiency since the tilt of collector magnetization $\mathbf{M^C}$ is proportional to the transverse component of total magnetic moment carried by the spin current pulse. A maximum is found for $d_E\approx7$~nm. 

Since the PSSW amplitude provides no information about the shape of spin current pulses we aim to measure it directly. The spin current can be obtained by detecting the second harmonic (SH) signal at 400~nm and calculating the SH contrast $\rho_{2\omega}$ as the ratio of difference and sum of SH intensities measured for up and down orientation of the emitter magnetization \cite{Alekhin2017}: 
\begin{equation}
    \label{eq:RhoSHmain}
    \begin{aligned}
\rho_{2\omega}(t) & =\frac{I_{2\omega}^{\uparrow}-I_{2\omega}^{\downarrow}}{I_{2\omega}^{\uparrow}+I_{2\omega}^{\downarrow}}= R^S(t)+R^C(t)= \\
& =R^S_j[\mathbf{j^S}(t)]+R^S_n[\mathbf{n^S}(t)]+R^C[\mathbf{M^C}(t)].
\end{aligned}
\end{equation}
Here, $R^C$ is the sum of SH responses of the two collector interfaces determined by the polarization geometry and $\mathbf{M^C}$ in the vicinity of interfaces. In equilibrium, $\mathbf{M^C}$ is directed along $x$-axis (Fig.~\ref{fig:Fig2}a
). Therefore, $M^C_y(t)$ which plays a role in the \emph{p}-in, \emph{p}-out geometry used here, is the transient component induced by the spin current $\mathbf{j^S}$. Thus, $R^C(t)$ reflects the PSSW dynamics in the collector while the spin transport in the spacer determines $R^S(t)$ consisting of two contributions: (i) the SH response of Au/Fe interface to the collector $R^S_n$ proportional to the local spin density $\mathbf{n^S}$ at the Au side and (ii) the bulk SH response of Au spacer proportional to $\mathbf{j^S}$ (see Supplemental Material \cite{supp}, Note 3). Owing to the similarity of $\mathbf{n^S}(t)$ and $\mathbf{j^S}(t)$ discussed later, $R^S(t)$ nearly resembles the shape of spin current pulse.

\begin{figure}
\centering
\includegraphics[scale=0.7]{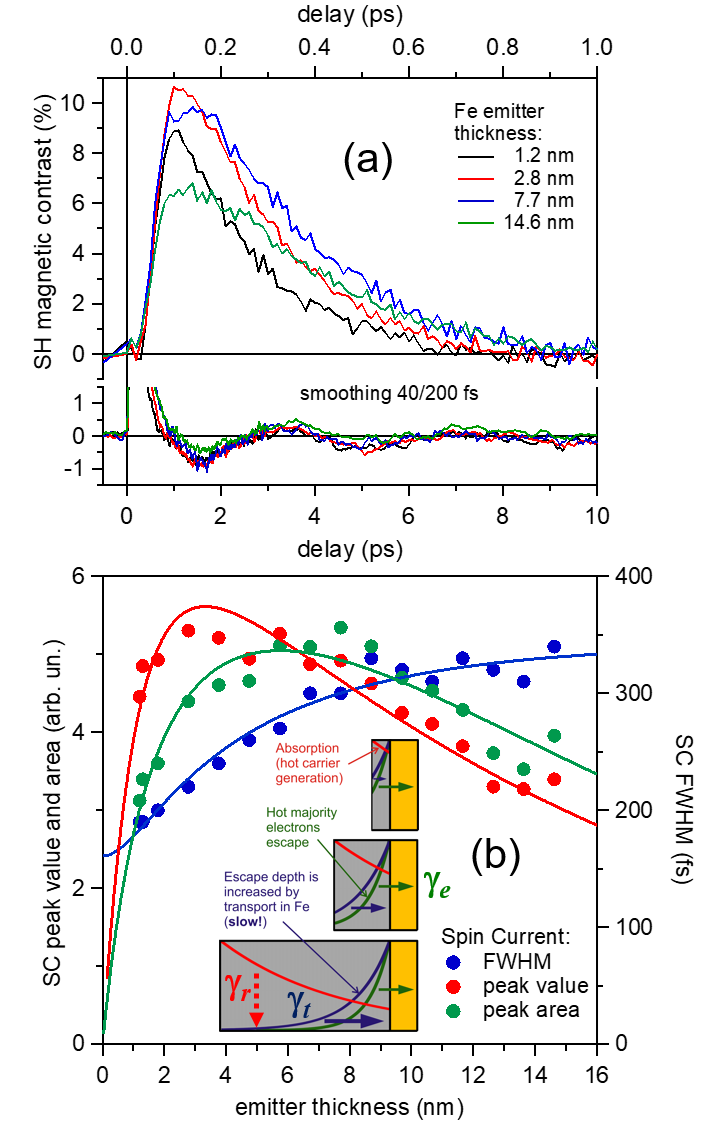}
\caption{(a) Exemplary traces of transient magnetic contrast of second harmonic signal, which is proportional to the spin current density, shown for indicated emitter thicknesses. (b) Spin current peak value, area, and FWHM vs. the emitter thickness. Solid curves are the fit to the model described in the text and illustrated in the inset.}
\label{fig:Fig3}
\end{figure}

Exemplary traces $\rho_{2\omega}(t)$ (Fig.~\ref{fig:Fig3}a) show a peak at $\sim$100~fs with a fast ($\sim$50~fs) rise time and a much longer ($\sim$300~fs) decay time (top). The latter increases with increasing $d_E$ by more than 50$\%$ while the peak amplitude changes with $d_E$ non-monotonously. This peak is followed by damped oscillations (bottom) at the frequency of first PSSW mode caused by $R^C$. Since their amplitude is about 10 times smaller than the peak amplitude, we conclude that the peak in Fig.~\ref{fig:Fig3}a represents the dynamics of spin current with sufficient accuracy.

Fitting the data in Fig.~\ref{fig:Fig3}a would require a microscopic model based on an approach like presented in \cite{Ritzmann2020}, which goes beyond the scope of this letter. Instead, we develop a simplified  model allowing us to identify the relevant processes occurring in the system.
For this, we characterize the observed dynamics in terms of effective parameters of the spin current pulse shape. For each emitter thickness $d_E$, we determine the observed peak value of traces in Fig.~\ref{fig:Fig3}a, the peak area, and the pulse duration given in terms of full width at half maximum (FWHM). Fig.~\ref{fig:Fig3}b summarizes these results. The largest peak value is observed at $d_E\approx3$~nm while the maximum of spin current peak area is shifted to $d_E\approx7$~nm owing to the increase of FWHM from $\sim$200~fs to $\sim$350~fs within the explored range of $d_E$. Remarkably, the dependence of the peak area on $d_E$ indeed perfectly matches that of the amplitude of 1st PSSW mode (see Fig.~\ref{fig:Fig2}b) as discussed above.

Such a behavior of spin current pulse parameters can be understood within an \emph{emission model} as illustrated in the inset of Fig.~\ref{fig:Fig3}b: The absorption of pump pulse in Fe (left, gray area) leads to a concentration of hot carriers $P(z)$ decreasing from the MgO/Fe ($z$=0) to the Fe/Au ($z=d_E$) interface as shown by the red curve. Subsequently, hot spin-up electrons are emitted within a relatively short time from the sub-interface volume of Fe layer with a characteristic emission depth $\lambda_e$ into the Au layer (right, yellow area). The emission probability is described by a function $f(\lambda_e,z)$ (green curve) and the peak amplitude is proportional to the total number of emitted electrons 
\begin{equation}
    \label{eq:F(d)}
F(d_E)=\int_0^{d_E}{P(z)f(z)dz}.
\end{equation}
Obviously, at small thicknesses one has $F(d_E)\propto d_E$. After a certain Fe thickness $F(d_E)$ starts to decrease with increasing $d_E$ together with the overlap of $P(z)$ and $f(z)$ (see Fig.~\ref{fig:Fig3}b, inset) in qualitative agreement with our observation. 

Following this "quasi-instantaneous" emission, the (super)diffusive transport in Fe delivers hot electrons from the depth of emitter to the Fe/Au interface. This leads to a weaker emission spread over larger time interval, causing the trailing edge of spin current pulse in Fig.~\ref{fig:Fig3}a. The effective depth of "emission region" on this longer time interval is enlarged as schematically shown by blue curve in the inset of Fig.~\ref{fig:Fig3}b. Correspondingly, the maximum of pulse area will shift towards larger $d_E$ and FWHM will increase, both in qualitative agreement with our observations. Formally, we can describe this "slow" contribution by the following function
\begin{equation}
    \label{eq:G(d)}
G(d_E)=\int_0^{d_E}{dz\int_0^{d_E}{[P(z)-f(z)]g(z,z')f(z')dz'}}
\end{equation}
with the effective transport probability $g(\kappa,\lambda_t,z,z')$. Here $\lambda_t$ is the characteristic length scale of hot spin-up electrons transport during their decay time $\tau_d=1/\gamma_d$. The decay rate $\gamma_d=\gamma_r+\gamma_t$ is given by the sum of hot carrier population relaxation rate $\gamma_r=1/\tau_r$ ($\tau_r$ corresponds to the hot electron lifetime in the absence of transport) and transport rate $\gamma_t$. Correspondingly, the transport efficiency is determined by $\kappa=\gamma_t/\gamma_d$.

The spin current peak area is proportional to $F+G$ and the effective pulse duration can be calculated as
\begin{equation}
    \label{eq:tau(d)}
\tau(d_E)=[\tau_F F(d_E)+\tau_G G(d_E)]/[F(d_E)+G(d_E)],
\end{equation}
where $\tau_F$ and $\tau_G$ are the effective durations of "fast" and "slow" contributions. For a quantitative description of observable parameters we introduce $P$, $f$, and $g$ functions of exponential shape (see Supplemental Material \cite{supp}, Note 4). The fit to experimental data (solid curves in Fig.~\ref{fig:Fig3}b) results in  $\lambda_t=4.3\pm0.9$~nm, $\lambda_e=2.1\pm0.1$~nm, $\tau_F=161\pm12$~fs, $\tau_G=532\pm33$~fs, and $\kappa=0.15\pm0.03$.

The obtained value of $\lambda_e$ is determined by the inelastic mean free path of hot majority electrons in Fe of $\sim$6~nm \cite{supp} in excellent agreement with earlier results \cite{Zhukov2006,Parkin2002}. The average velocity of electron transport along $z$ can be estimated from $\lambda_t$ and $\tau_G$ to be $\sim$8~nm/ps, which is a typical value for electron diffusion in metals \cite{Hohlfeld2000} and two orders of magnitude smaller than the ballistic electron velocity ($\sim$500~nm/ps) \cite{Zhukov2006}. This indicates an essentially diffusive character of the transport and justifies our model. $\tau_G$ is associated with the decay rate of hot electron population $\gamma_d\approx$2~ps$^{-1}$ resulting in a transport rate of $\gamma_t=\kappa\gamma_d\approx0.3$~ps$^{-1}$ and a relaxation rate of $\gamma_r=\gamma_d-\gamma_t\approx1.7$~ps$^{-1}$ (see Fig.~\ref{fig:Fig3}b, inset). The latter corresponds to the relaxation time of the hot carrier population in the absence of transport $\tau_r\approx600$~fs, which is close to the 500~fs electron-electron thermalization time reported e.g. for Ni \cite{Rhie2003} and a factor of two smaller than the electron-phonon equilibration time observed for an isolated Fe film where electron transport effects are excluded \cite{Razdolski2017a,Alekhin2016PhD}. 

For \emph{primary} excited hot majority electrons one would expect a characteristic emission time of a few fs due to the large transmittance at the Fe/Au interface \cite{Alekhin2017}. This is much less than $\tau_F$ obtained in the fit to our experiments. There are two obvious reasons for that: (i) the spin current pulse can broaden in the spacer due to the distribution of directions of electron emission and scattering in Au, (ii) the generation of \emph{secondary} hot majority electrons during the decay of hot holes and minority electrons left in Fe. To explore the roles of these factors we describe the behavior of the SH contrast using a simple model of superdiffusive spin transport in Au. We consider only the response for the thinnest emitter (1.2 nm), where the diffusive transport in Fe to a good extend can be neglected, simplifying the analysis. 

We assume an isotropic electron emission and energy- and direction-independent ballistic velocity of $v_b=$1.2~nm/fs \cite{Zhukov2006}. We also consider an effective energy-independent lifetime $\tau_k=$50~fs for the momentum scattering in Au including both electron-electron and electron-lattice processes. Unlike the hot electron density which can be increased by secondary carrier excitation in Au (this is important e.g. for the photoemission detection \cite{Beyazit2020}), the spin density is conserved in electron-electron interactions. However, it can be reduced by electron-lattice spin relaxation on the time scale $\tau_s$ set to 500~fs. Assuming isotropic scattering and fast screening of hot electron charge by electrons at the Fermi level, we obtain a simple Boltzmann equation as described in Note 5 of Supplemental Material \cite{supp}. A numerical solution of this equation provides the spin current $j^S_{z,y}(z,t)$ and spin density $n^S_y(z,t)$ generated by instantly emitted electrons \endnote{In the \emph{p}-in, \emph{p}-out geometry, $j^S_{z,y}$, the component with the velocity along $z$ and the spin polarization along $y$, and $n^S_y$ are important.}. Convolving them with a Gaussian of 20~fs width to account for the experimental time resolution, we obtain the spin current $j^S_{prim}(z,t)$ and spin density $n^S_{prim}(z,t)$ generated by \emph{primary} electrons. The spin density at the Au/Fe interface is proportional to $R^S_{n,prim}(t)$ in Eq.~(\ref{eq:RhoSHmain}) while $R^S_{j,prim}(t)$ is calculated by integrating the product of $j^S_{prim}(z,t)$ and response function over $z$ (see Supplemental Material \cite{supp}, Note 5). The response function accounts for retardation and absorption of fundamental and SH waves in Au. The contribution of \emph{secondary} majority electrons is calculated as follows
\begin{equation}
    \label{eq:RsecMain}
R^S_{sec}(t)=\alpha_{sec}\int_0^t{S^e_{sec}(\tau_g,\tau_e,t') R^S_{prim}(t-t')dt'},
\end{equation}
where $\alpha_{sec}$ is the ratio of emitted secondary and primary electrons. The normalized emission source function $S^e_{sec}$ has two characteristic timescales: (i) the rise time $\tau_g\approx5$~fs determined by a typical lifetime of holes and minority electrons in Fe  \cite{Zhukov2006} and (ii) the decay time $\tau_e$ predominantly determined by the electron emission rate $\gamma_e$ as $\tau_e^{-1}=\gamma_e+\gamma_r$. Since the exact calculation of $S^e_{sec}$ goes beyond this work, we approximate it by a phenomenological double-exponential expression (see Supplemental Material \cite{supp}, Note 5).

\begin{figure}
\centering
\includegraphics[scale=0.6]{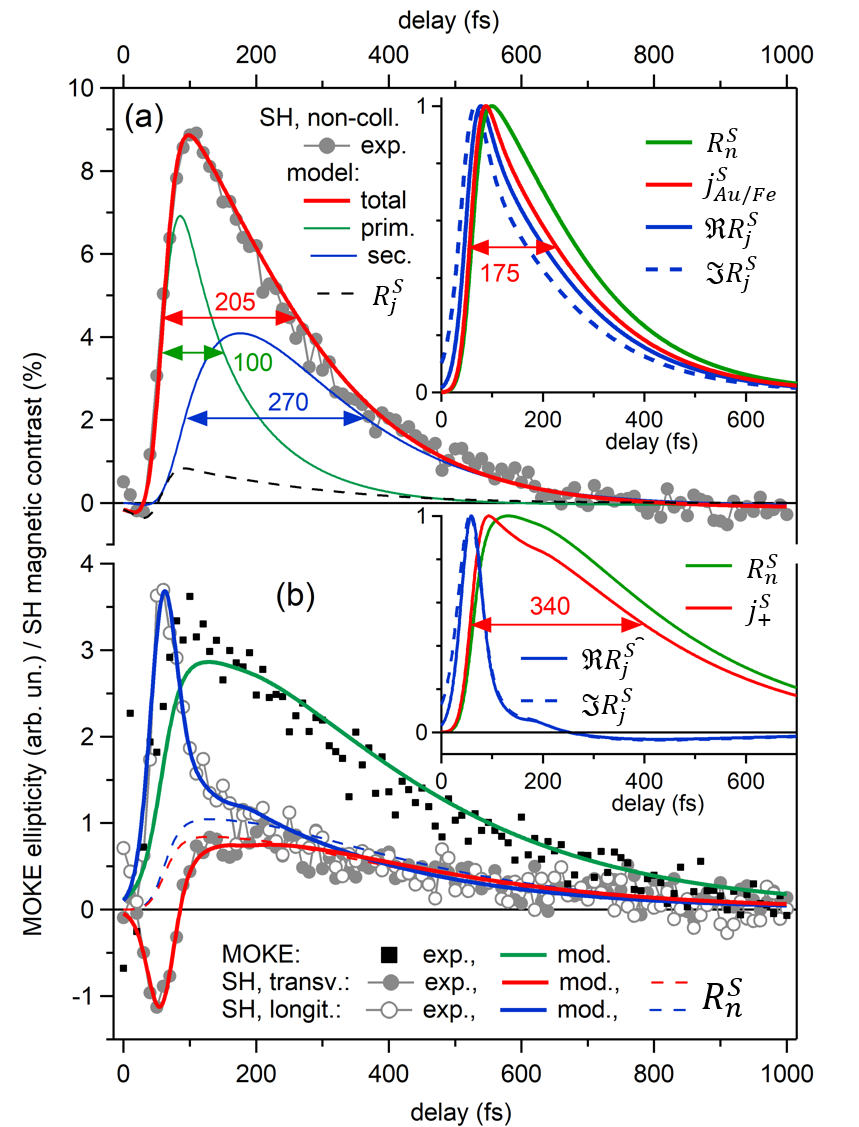}
\caption{Transient magnetic contrast of second harmonic signal (a, b) and MOKE ellipticity (b) measured for 1.2~nm-thick emitter vs. pump-probe delay in Fe/Au/Fe (a) and Fe/Au (b) structures in non-collinear (a), transversal, and longitudinal (b) magnetic configurations. Curves are calculated within the model for superdiffusive spin transport described in the text. Insets show the shapes of selected contributions (see text). Arrows with numbers indicate the FWHM in femtoseconds.}
\label{fig:Fig4}
\end{figure}

The result of calculations assuming total absorption of spin by the collector \cite{Alekhin2017} is shown in Fig.~\ref{fig:Fig4}a with the red curve. Green and blue curves show particular contributions of primary and secondary electrons. The inset compares normalized spin current $j^S_{z,y}$ at the interface to the collector (red) with spin-density- and spin-current-induced contributions to $\rho_{2\omega}$ [cf. Eq.~(\ref{eq:RhoSHmain})]:  
they are quite similar although $R^S_n$ is somewhat broader while $R^S_j$ is a bit narrower and shifted to smaller delays (see Supplemental Material \cite{supp}, Note 6).
\endnote{Since the relative phase of spin-current-induced susceptibility is \emph{a priori} unknown, we have to consider both complex components, $\Re R^S_j$ and $\Im R^S_j$, on equal footing (see Supplemental Material \cite{supp}, Note 6).}. 
This similarity indicates that the transport character is not far from the ballistic limit, where the electron velocity distribution does not change much with time and $j^S_{z,y}(z,t)$ is mostly determined by $n^S_y(z,t)$. 

For the given collector thickness, $R^S_n$ dominates the response while $R^S_j$ (dashed curve in Fig.~\ref{fig:Fig4}a) is probably out of phase with respect to the leading (non-magnetic) interface contribution and $R^C$ only provides a little offset at large delays. To increase the robustness of our analysis, we add the data shown in Fig.~\ref{fig:Fig4}b, which are obtained at exactly the same structure as in Fig.~\ref{fig:Fig4}a, however without the collector \endnote{For such Fe/Au bilayers, we have evidences of importance of both $R^S_j$ and $R^S_n$ contributions \cite{Alekhin2019}.}. All model parameters describing the spin transport are identical. 
Note that the inset in Fig.~\ref{fig:Fig4}b shows the same quantities as the inset in Fig.~\ref{fig:Fig4}a with the following important difference: The red curve in the inset of Fig.~\ref{fig:Fig4}b shows the incident current $j^S_+$ since in this case the total reflection of electrons leads to $j_S$=0 at the Au surface. 
$R^S_n(t)$ is again similar to $j^S_+(t)$, however $R^S_j(t)$ differs dramatically due to the reflected spin current. The model curves fit the data well (Fig.~\ref{fig:Fig4}b) with a positive $R^S_n$ term (dashed curves) and $\Re R^S_j$ contribution taken positive (negative) for the longitudinal (transversal) magnetic configuration \cite{Alekhin2019}. Moreover, the integral response of $n^S(z,t)$ calculated similar to $R^S_j$ (green curve) describes the transient MOKE signal \endnote{The deviations probably can be explained by residual contribution from the emitter and pump-induced variations of magneto-optical constants \cite{Razdolski2017a}.}.

The experimental data are fit by setting $\tau_e=130$~fs in Eq.~(\ref{eq:RsecMain}) providing an emission efficiency of $\gamma_e/(\gamma_e+\gamma_r)\approx0.8$. This means that in thin Fe emitters about 80$\%$ of generated secondary carriers are converted into hot majority electrons and emitted into Au. The data are consistent with $\alpha_{sec} =1.4 $, the ratio determining relative amplitudes of green and blue traces in Fig.~\ref{fig:Fig4}a, showing that carrier multiplication in Fe increases the injected spin moment by a factor of 2.4. However, the spin relaxation in the 70~nm-thick spacer reduces the density of spin-up electrons delivered to the collector to 115$\%$ of the density of optically excited majority electrons. The latter can be estimated  to $2.8\pm0.3$~nm$^{-2}$ from the absorbed fluence taking into account that only 40$\%$ of carriers are excited in the majority sub-band \cite{Melnikov2011}. The magnetic moment absorbed by the collector is measured by the amplitude of MOKE signal following Ref.~\cite{Razdolski2017} and gives $3.1\pm0.2~\mu_B$/nm$^2$. This agreement is an independent validation of the model. Therefore, we conclude that the developed approach is well justified and promising for the determination of key material parameters (such as carrier momentum and spin scattering rates, velocity, interface transmittance, etc.) from magneto-optical experiments by fitting extended data sets based on systematic variation of layer thicknesses.       

Summarizing, using the high sensitivity of nonlinear-magneto-optical probe to spin states of interfaces and transient bulk inversion symmetry breaking by spin currents,  we have demonstrated the control of ultrashort spin current pulse shape by varying the thickness of spin emitter in Fe/Au/Fe epitaxial structures. The developed simple phenomenological model allows to identify and measure the underlying material parameters.
Based on the assumption of superdiffusive spin transport in Au spacer described in the framework of Boltzmann theory, we have achieved a consistent description of experimental data in both Fe/Au/Fe and Fe/Au structures, confirmed a nearly-ballistic transport regime in Au, and shown the importance of carrier multiplication in the emitter.

The authors thank I.~Razdolski, T.~Kampfrath, and  M.~Weinelt for fruitful discussions.  Funding by the German research foundation (DFG) through CRC/TRR 227 (projects B01 and B04) is gratefully acknowledged.

\end{document}